\documentclass[journal=jpcbfk,manuscript=article]{achemso}
\usepackage[version=3]{mhchem}

\usepackage{graphicx}  
\usepackage{dcolumn}   
\usepackage{bm}        
\usepackage{amssymb}   
\usepackage{color}
\usepackage{epsfig}
\usepackage{epstopdf}
\SectionNumbersOn

\author{Anatolii V. Mokshin}
\email{anatolii.mokshin@mail.ru}
\affiliation[Kazan (Volga region)Federal University]{Department of Computational Physics,
	Kazan (Volga region) Federal University, Kazan 420008, Russia}

\author{Roman V. Vlasov}
\affiliation[Kazan (Volga region) Federal University]{Department of Computational Physics,
	Kazan (Volga region) Federal University, Kazan 420008, Russia}

\title[\texttt{achemso}]
{Liquid-Liquid Crossover in Water Model:\\ Local Structure \textit{vs.} Kinetics of Hydrogen Bonds}

\begin{document}
	
	\begin{tocentry}
		\includegraphics{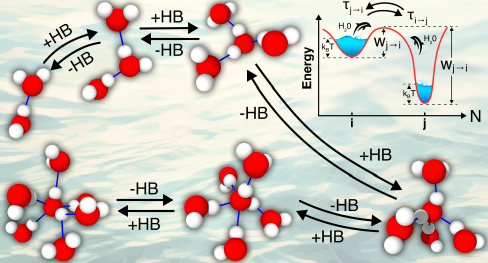}	
	\end{tocentry}

\begin{abstract}
    In equilibrium and supercooled liquids, polymorphism is manifested by thermodynamic regions defined in the phase diagram, which are predominantly of different short- and medium-range order (local structure). It is found that on the phase diagram of the water model, the thermodynamic region corresponding to the equilibrium liquid phase is divided by a line of the smooth liquid-liquid crossover.
    In the case of the water model, this crossover is revealed by various local order parameters and corresponds to pressures of the order of $3\,150 \pm 350$~atm at ambient temperature. In the vicinity of the crossover, the dynamics of water molecules change significantly, which is reflected, in particular, in the fact that the self-diffusion coefficient reaches its maximum values. In addition, changes in the structure also manifest themselves in changes in the kinetics of hydrogen bonding, which is captured by values of such the quantities as the average lifetime of hydrogen bonding, the average lifetimes of different local coordination numbers, and the frequencies of changes in different local coordination numbers. An interpretation of the hydrogen bond kinetics in terms of the free energy landscape concept in the space of possible coordination numbers is proposed.
\end{abstract}

\section{Introduction}\label{sec: Introduction}
	
Crystalline solids are characterized by polymorphism: equilibrium phases with different structures are possible. If in the case of crystals and quasicrystals the term ``\textit{structure}'' implies a certain regularity in the arrangement of the particles (atoms, molecules or ions) that form them, in the case of liquids the term ``\textit{structure}'' implies a statistically averaged configuration that characterizes the mutual arrangement of the particles. Thus, even in the presence of high particle mobility and their significant displacements relative to each other, typical of classical equilibrium liquids at finite temperatures, a statistically averaged configuration remains unchanged. As found for many single-component liquids~\cite{Brazhkin/Lyapin_PCM_2003,Tanaka_JCP_2020}, the thermodynamic region in their phase diagrams corresponding to both equilibrium and supercooled liquid phases is divided into subregions of ``low density liquid'' (LDL) and ``high density liquid'' (HDL) states.
The structure changes significantly at the transition between these states, known as the liquid-liquid transition (LLT).  This transition is most evident in atomistic and molecular liquids, where the interparticle interaction is essentially non-spherical  and/or promotes the formation of network structures. The LLT is found in metallic melts of cerium~\cite{Cerium} and bismuth~\cite{Bismuth_1,Bismuth_2}, in pure silicon~\cite{Silicon_1, Silicon_2,Silicon_3}, phosphorus~\cite{Phosphorus_1,Phosphorus_2}, sulfur~\cite{Sulfur} as well as in melts of triphenyl phosphite~\cite{TPP_1,TPP_2}, germanium oxide~\cite{Brazhkin/Lyapin_PCM_2003} and boron oxide~\cite{B2O3}.
	
In the case of water, a discontinuous LLT, which has features of a first-order phase transition, appears for supercooled states~\cite{Mishima_1984,Mishima_1985,Exp_LLT_1,Exp_LLT_2,Exp_LLT_3}.  The LLT line crosses the region of deep supercooling, separating the low-density amorphous (LDA) ice phase  and the high-density amorphous (HDA) ice phase, a region of moderate supercooling, separating the LDL and HDL states, and presumably ends at the so-called second critical point (LLCP) with the density $\rho_{LLT}^{(c)}$, the temperature $T_{LLT}^{(c)}$ and the pressure $p_{LLT}^{(c)}$.
At present, there is no known an exact analytical equation $f(p_{LLT},T_{LLT})=0$ analogous to the Clausius-Clapeyron equation, derived from thermodynamic considerations, that uniquely defines the transition line and the corresponding critical point, i.e., values of the pressure $p_{LLT}^{(c)}$ and the temperature $T_{LLT}^{(c)}$.
On the other hand, the known experimental measurements~\cite{Exp_LLT_1,Exp_LLT_2,Exp_LLT_3} report different results, which, in turn, differ from the results of \textit{ab initio} molecular dynamics simulations~\cite{Gartner_III_PNAS_2020,Gartner_III_PRL_2022}.   Classical molecular dynamics simulations with various model potentials quite expectedly yield unique results for the LLT~\cite{Poole_Nature_1992,crit_point_TIP4P_2005,Yicun_JChemPhys_2016,Debenedetti_Science_2020}. Thus, at this point,
one can speak of a region on the $(p,\;T)$ phase diagram of water, where the LLT is likely to be observed.

\begin{figure}[t!]
		\centering
		\includegraphics[width=1\linewidth]{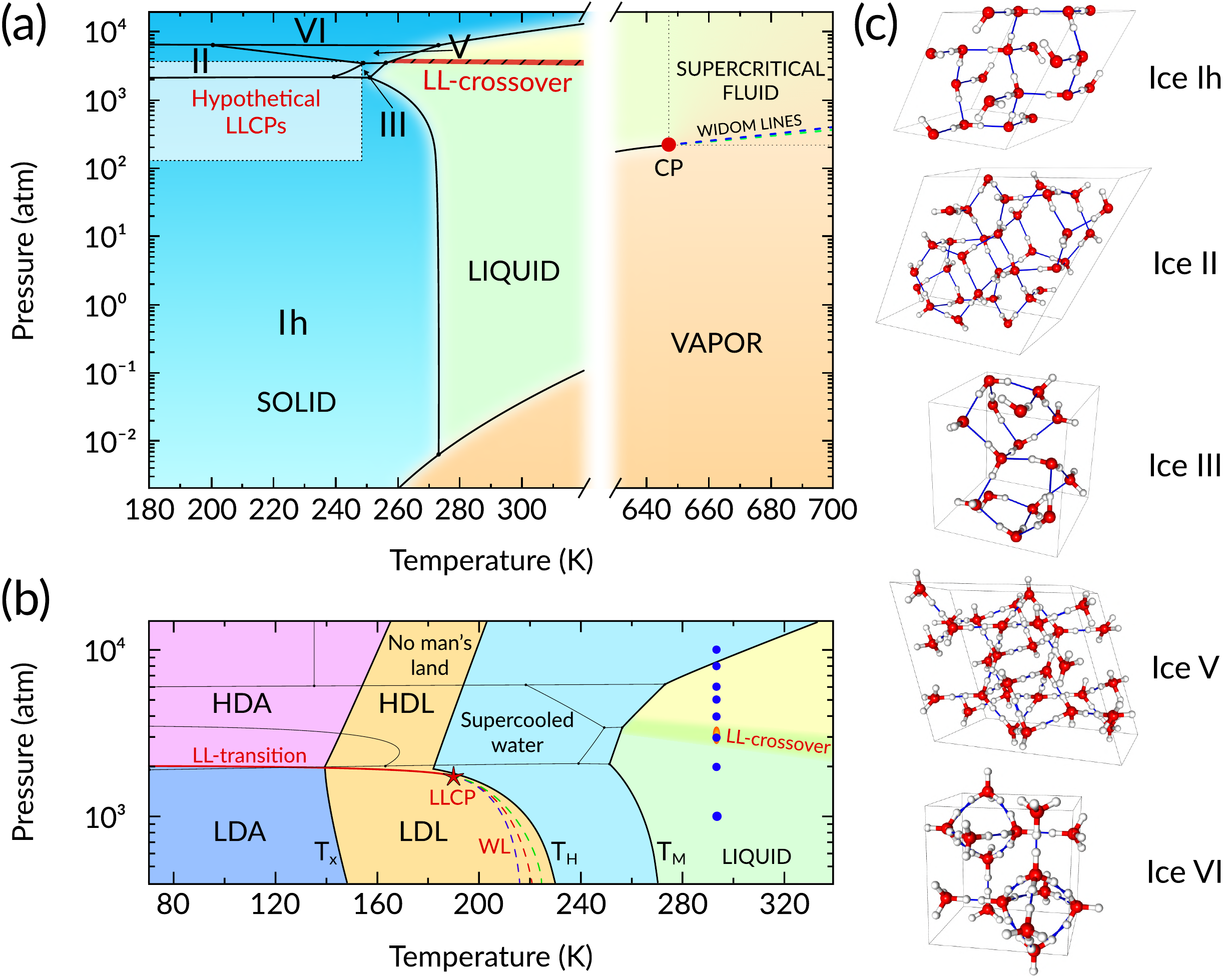}
		\caption{(Color online)
(a) Phase diagram of water for the wide range of pressures $p$ and temperatures $T$. The highlighted rectangular region of hypothetical LLCPs contains the currently known LLCP values~\cite{Mishima_book_2021,Poole_Nature_1992,Yamada_PRL_2002,Debenedetti_PhysToday_2003,Poole_JPCM_2005,crit_point_TIP4P_Ew,Paschek_PRL_2005,Liu_JChemPhys_2009,crit_point_TIP4P,crit_point_TIP4P_2005,Cuthbertson_PRL_2011,Bertrand_2011,Holten_JChemPhys_2012,Holten_SciRep_2012,Yaping_PNAS_2013,Yicun_JChemPhys_2016,Caupin_JChemPhys_2019,Debenedetti_Science_2020,Gartner_III_PNAS_2020,Gartner_III_PRL_2022,Mishima_JPCB_2023}; the thick red line denotes the hypothetical LL-crossover line~\cite{Saitta}; the red solid circle labeled (CP) denotes the critical point where the saturation line ends and from which the Widom lines (blue and green dashed lines) originate~\cite{Gallo_Widom_lines}. Plotted on the basis of data from Refs. \citenum{Chaplin_2019, Petrenko_1999}.
\\
		(b) Fragment of ($p,\;T$)-phase diagram containing the hypothetical LL-transition line with LLCP as well as the Widom lines coming from this point (according to Ref.~\citenum{Mallamace_LLT}); $T_x$ is the crystallization temperature, $T_H$ is the homogeneous crystal nucleation temperature and $T_m$ is the melting temperature. The blue dots on the isotherm $T=293$~K denote the states considered in this paper; the red segment denotes the region of smooth LL-crossover.   \\
        (c) Ice diagrams for Ih, II, III, V and VI crystalline phases.
        }
		\label{fig: phas_diagram}
	\end{figure}

The available experimental and simulation results reveal the following common features (see Figure~\ref{fig: phas_diagram}):
	
	\noindent (i) On the ($p,\;T$)-phase diagram, the LLT line is characterized by a small \textit{negative} slope relative to the temperature axis, and this transition in water is induced by pressure from the range [$1\,000;\, 3\,000$]~atm~\cite{Gromnitskaya_PRB_2001,Tale_of_Two_Liquids_review,Mishima_book_2021}.
	
	\noindent (ii) The critical point -- LLCP -- is assumed to be in the temperature region bounded by the crystallization temperature $T_x$ and the melting temperature $T_m$.
	
	\noindent (iii) The currently known LLCP values are in the temperature range $T \in [180;\, 247]$\;K and pressure range $p \in [130;\, 3\,400]$~atm (based on data from Refs.~\citenum{Mishima_book_2021,Poole_Nature_1992,Yamada_PRL_2002,Debenedetti_PhysToday_2003,Poole_JPCM_2005,crit_point_TIP4P_Ew,Paschek_PRL_2005,Liu_JChemPhys_2009,crit_point_TIP4P,crit_point_TIP4P_2005,Cuthbertson_PRL_2011,Bertrand_2011,Holten_JChemPhys_2012,Holten_SciRep_2012,Yaping_PNAS_2013,Yicun_JChemPhys_2016,Caupin_JChemPhys_2019,Debenedetti_Science_2020,Gartner_III_PNAS_2020,Gartner_III_PRL_2022,Mishima_JPCB_2023}).
	
	\noindent (iv) The LLCP is located near the isobar, which contains a ternary point for hexagonal crystalline ice (ice-Ih), tetragonal crystalline ice (ice-III) and equilibrium water phases. Near this isobar, the water-ice coexistence line changes the slope from negative to positive.
	
	\noindent (v) For the local structure of the LDL state, the characteristic interparticle distances and angles in the triplets of neighboring molecules correlate with the crystal lattice constants of tetragonal and rhombohedral ice, indicating a high degree of tetrahedricity~\cite{PNAS_Francesco_Mallamace_2009}. The HDL state arises due to the densest packing of water molecules, where the directional bonds, that are typical of water and are responsible for the formation of the tetrahedral structure, appear much weaker and practically do not determine the character of the local order.
	
	In the overcritical region at temperatures $T > T_{LLT}^{(c)}$, the thermodynamic response functions -- the isobaric heat capacity $C_p$, the isothermal compressibility $\beta_T$, the thermal expansion coefficient $\alpha_p$ -- reveal extremes that form the corresponding lines. In the vicinity of the LLCP, these lines merge into the so-called Widom line and converge to the LLCP~\cite{Tale_of_Two_Liquids_review}. In turn, according to the original definition~\cite{Stanley_PNAS_2005}, the Widom line is a line originating from a critical point and defined by the ($p,\;T$) points in the phase diagram at which the correlation length takes maximum values. Thus, it is assumed that there should be at least two Widom lines in the phase diagram of water, one referring to the supercritical fluid and coming from the critical point ($\rho_c,\; p_c,\;T_c$), and the other referring to the LLT and coming from the LLCP ($\rho_{LLT}^{(c)},\;p_{LLT}^{(c)},\;T_{LLT}^{(c)}$) [see Figure~\ref{fig: phas_diagram}].

In addition, in the specific overcritical region at temperatures $T > T_{LLT}^{(c)}$, there are also two types of local structures corresponding to LDL and HDL states, where the concentration ratio of these structures with temperature and pressure changes smoothly. Thus, the phase diagram also exhibits \textit{a smooth LL-crossover line}, originating presumably from the LLCP and continuing to higher temperatures and defining subregions in this phase diagram where either LDL- or HDL-local structures predominate. At the crossover, a discontinuous change will be revealed only for some local structural characteristics, whereas all macroscopic and thermodynamic parameters will change smoothly~\cite{Khusnutdinoff_Mokshin_JNCS_2011,Khusnutdinoff_Mokshin_PhysA_2012}. In fact, this crossover line is similar to the so-called Frenkel line, which, in turn, on the phase diagram of a supercritical fluid divides the regions of predominance of oscillatory or diffusive dynamics of molecules~\cite{Frenkel_line_PRE2012,Frenkel_line_PRL2013}.

The aim of the present study is to clarify how changes in the local structure associated with the LL-crossover are manifested in the mobility of water molecules as well as in the kinetics of hydrogen bond (HB) formation. Using a water model as an example, the LDL and HDL states are considered for the isotherm corresponding to ambient temperature and the key structural, transport and kinetic properties for these states are determined. The main focus is on how the crossover is reflected in such the properties as the average HB lifetime, the average lifetimes of different local configurations with the coordination numbers $\mathcal{N}=1$, $2$, $\ldots$, $6$, and the rates of change of these local configurations. The obtained results allow one to provide unique information about the changes in the thermodynamics of HB formation that occur in the vicinity of the LL-crossover.

\section{Methods} \label{sec: Methods}

\subsection{Simulation Details}

For the purposes of this study, it is not necessary that the water model under consideration reproduce as accurately as possible all the physical properties of real water. A necessary condition for choice of a model is the presence of bonds in the effective interparticle interaction, which are capable of forming a network of HBs as in water. In addition, a model should be relatively simple for simulations, so that sufficiently large time scales can be covered and different states can be considered. The non-polarizable water models TIP4P-Ew and TIP4P/2005 reproduce the density over a range of temperatures, as well as the density maximum $\rho_m(T)$, approximating the actual values of temperature $T$ and density $\rho_m$ for water~\cite{TIP4P-2005_1,TIP4P-2005_2}. These models reproduce the features of the melting line trend of water over a wide range of pressures. In contrast, the TIP4P/2005 model produces more correct values for thermal coefficients (isothermal compressibility, coefficient of thermal expansion) and caloric coefficients (e.g. isobaric heat capacity)~\cite{TIP4P-2005_1,compressibility_supercool}. Although this study is concerned with equilibrium liquid states, it is important to note that the TIP4P/2005 model produces a large number of intrinsic crystalline water phases over a wide pressure range~\cite{TIP4P-2005_1,phase_trans_HB_network}. Furthermore, the TIP4P/2005 model gives a better agreement with experimental viscosity data in the temperature range from 273~K to 293~K compared to other non-polarizable potentials: SCP/E, TIP4P and TIP4P-Ew~\cite{Wikfeldt_2011,malinovsky2015origin,Bird2002,Markesteijn2012}. Thus, the TIP4P/2005 model is one of the most accurate classical non-polarizable liquid water models~\cite{Singh_2016}.

The molecular dynamics simulations with the LAMMPS software GPU package were performed for $N=4\,096$~molecules enclosed in a cubic box with periodic boundary conditions and  interacted via the TIP4P/$2005$ potential~\cite{TIP4P-2005_1,LAMMPS,Brown11,Brown12,Brown13,Trung15,Trung17,Nikolskiy19}. The isothermal-isobaric ensemble was realized by means of the Nos\'{e}-Hoover thermostat and barostat with the relaxation constants~$\tau_T=0.1$~ps and $\tau_p=1.0$~ps, respectively~\cite{Nose_NpT}. The cutoff radii for the Coulomb and Lennard-Jones interactions were taken as $r_{c,Coul}=r_{c,LJ}=10$~\AA. The long-range electrostatic interactions were treated using the Particle–Particle–Particle-Mesh (PPPM) algorithm with a splitting factor of $0.311$~\AA$^{-1}$ and a grid of $45 \times 45 \times 45$~\cite{PPPM}. The bond lengths and angles in the rigid water molecule were controlled by the SHAKE algorithm~\cite{SHAKE}. The PPPM and SHAKE tolerances were set to $1.0 \cdot 10^{-5}$. Integration of the equations of motion was performed with the time step~$\Delta\tau = 1.0$~fs.

The study covers the thermodynamic states along the isotherm $T=293$~K at pressures from the range~$p \in [1.0;\, 10\,000]$~atm. All the states correspond to the equilibrium liquid phase. Each simulation configuration was initially equilibrated for the time~$t_{eq}=1.0$~ns. To calculate the physical properties, the molecular dynamics simulations were performed over the time window $t=9.0$~ns.

\subsection{Main characteristics}

For each thermodynamic state considered, the following characteristics are determined.

The radial distribution function $g(r)$ carries information about a structure of the system under consideration. This function is associated with the probability of finding two arbitrary particles at a distance $r=|\vec{r}|$ from each other and can be defined as follows
\begin{equation}\label{eq: g(r)}
	g(r) = \lim_{dr \to 0} \frac{V n(r)}{4 \pi r^2 N_{\mathrm{pairs}}} .
\end{equation}
Here, $n(r)$ is the average number of particle pairs located at a distance between $r$ and $r+dr$, $V$ is the volume of the system, and $N_{\mathrm{pairs}}$ is the number of unique pairs of particles~\cite{Levine2011,fairushin2020direct}.
The pronounced maxima in this function are located at distances indicating the most probable distances between the particles. The center of mass of a water molecule practically coincides with the center of mass of an oxygen atom of this molecule. So, it is convenient to characterize the structure of water by means of the radial distribution function of molecules determined by oxygen atoms, i.e. $g_{OO}(r)$.

It is convenient to take into account the local structural order associated with the nearest neighborhood of the particles by such a scalar quantity as the Wendt-Abraham parameter~\cite{Wendt_Abraham}
\begin{equation}\label{eq: r_WA}
	\tilde{r}=g(r_{min})/g(r_{max}),
\end{equation}
where $r_{min}$ and $r_{max}$ are the positions of the first minimum and maximum in the radial distribution function $g(r)$, respectively [see inset in Figure~\ref{fig: struct_parameters}(c)]. The Wendt-Abraham parameter can take values from the range [$0,\;1$].  In the case of a perfect crystal lattice, this parameter takes the value $\tilde{r}=0$; in the case of a gas, we have $\tilde{r}=1$. Thus, values of the parameter close to zero indicate a local structure close to crystalline.

For the case of the molecular system with the directional bonds as in water, the orientational ordering can be characterized by the average HB angle~$\langle \beta_{OOH} \rangle$, the tetrahedral order parameter $\langle Q \rangle$ and the orientational order parameter $\langle Q_6 \rangle$. The HB is defined according to geometric considerations. It is assumed that a pair of neighboring water molecules forms a HB if the relative distances $R_{OO}$ and $R_{OH}$ as well as the so-called HB angle $\beta_{OOH}$ do not exceed the values $R_{OO}^{(c)}=g_{OO}(r_{min})$, $R_{OH}^{(c)}=g_{OH}(r_{min})$ and $\beta_{OOH}^{(c)}=30^\circ$, respectively~\cite{Luzar_1993} [see inset in Figure~\ref{fig: struct_parameters}(d)]. Thus, average HB angle~$\langle \beta_{OOH} \rangle$ is defined as follows
\begin{equation}
    \langle \beta_{OOH} \rangle = \frac{1}{N S} \sum_{i=1}^{N} \sum_{j=1}^{S} \beta_{OOH}^{(i,j)},
\end{equation}
where $S$ is the instantaneous number of the HBs in which the $i$th molecule participates.
The tetrahedral order parameter $\langle Q \rangle$ evaluates the degree of tetrahedrality in the nearest neighborhood of water molecules and is defined as~\cite{Errington_Q_order_parameter}
\begin{equation}\label{eq: Q_order}
	\langle Q \rangle = \left\langle 1 - \frac{3}{8}\sum_{i=1}^{3} \sum_{j=i+1}^{4} \left(cos\theta_{ij}+\frac{1}{3}\right)^2 \right\rangle .
\end{equation}
Here, $\theta_{ij}$ is the angle formed by some molecule and its neighboring molecules $i$ and $j$. One has $\langle Q \rangle = 1$ for a perfect tetrahedral order, while for a random local arrangement of molecules it is $\langle Q \rangle = 0$. Angle brackets $\langle ... \rangle$ for this quantity and others below denote ensemble and time averaging.

The global orientational order parameter can be defined as follows~\cite{Q6_parameter,Q6_parameter_AVM,mokshin2009shear,Q6_2010_PRE}
\begin{equation}\label{eq: Q6}
    \langle Q_6 \rangle = \left \langle \left(  \frac{4\pi}{13} \sum_{m=-6}^{6}  \left| \frac{\sum_{i=1}^{N} \sum_{j=1}^{N_b(i)} Y_{6m}(\theta_{ij},\phi_{ij})}{\sum_{i=1}^{N}N_b(i)} \right|^2 \right)^{\frac{1}{2}} \right \rangle,
\end{equation}
where $Y_{6m}(\theta_{ij},\phi_{ij})$ are the spherical harmonics, $\theta_{ij}$ and $\phi_{ij}$ are the polar and azimuthal angles formed by the radius-vector $\vec{r}_{ij}$ and some reference system. Then, $N_b(i)$ denotes the number of nearest neighbours of molecule $i$ that are at a distance $|\vec{r}_{ij}|$ not exceeding $r_{min}$, i.e. $|\vec{r}_{ij}|<r_{min}$, where $r_{min}$ corresponds to the first minimum in the radial distribution function $g_{OO}(r)$. For a fully disordered system one has $\langle Q_6 \rangle \to 0$, whereas for perfect FCC and HCP crystalline phases it takes values $0.575$ and $0.485$, respectively~\cite{Q6_parameter_AVM}.

Based on the time-dependent configurations obtained from molecular dynamics simulations, we can estimate the self-diffusion coefficient as the slope of the mean-square displacement of a particle with respect to time $t$~\cite{Mokshin_PRE_2003,AVM_PRL_2005}:
\begin{equation}\label{eq: D_self}
	D_s = \frac{1}{6} \lim_{t \to \infty} \frac{d}{dt} \langle | \Delta \vec{r}(t)|^2 \rangle.
\end{equation}
The average HB lifetime $\langle \tau_{HB} \rangle$ can be evaluated with different definitions.

\noindent (a) First, the quantity $\langle \tau_{HB} \rangle$ is directly determined from the simulation results as the average bonding time of pairs of molecules, which is corrected for possible `false' HBs existing at times less than $0.2$~ps and corresponding to the librational dynamics of molecules.

\noindent (b) If the instantaneous average value of the number $\langle N_{HB} \rangle$ of the HBs and the total number $N_{all}$ of the HBs registered in the system for a time interval $t_{sim}$ are known, the quantity $\langle \tau_{HB} \rangle$ is defined as~\cite{Luzar2000}
\begin{equation}\label{eq: tau_HB_2}
	\langle \tau_{HB} \rangle  = \frac{\langle N_{HB}\rangle}{N_{all}}t_{sim}.
\end{equation}

\noindent (c) And, finally, the quantity $\langle \tau_{HB} \rangle$ appears as a parameter in the kinetic model for the reaction flux correlation function
\begin{equation} \label{eq: tau_HB_3}
    -\frac{dC_{HB}(t)}{dt} = \langle \tau_{HB} \rangle^{-1} C_{HB}(t)- k_2\, n(t).
\end{equation}
Here, $k_2$ is the breaking rate constant, and
\begin{equation}\label{eq: C_HB}
	C_{HB}(t) = \frac{\langle h(t)h(0) \rangle}{\langle h \rangle}
\end{equation}
is the HB autocorrelation function. The dynamical variable $h(t)$ equals unity, if a pair of molecules is bonded, and is zero otherwise~\cite{Luzar_1993, Luzar_1996, Luzar2000}. Further, $n(t)$ is the HB breaking function defined as
\begin{equation}\label{eq: n_t}
	n(t) = \int_{0}^{t} k_{in}(t')\,dt',
\end{equation}
and
\begin{equation}\label{eq: kin_t}
	k_{in}(t) = -\frac{\langle \dot{h}(0)[1-h(t)]H(t) \rangle}{\langle h \rangle}
\end{equation}
is the restrictive reactive flux function with
\begin{equation}\label{eq: k_t}
	\begin{matrix}
		H(t) =
		\left\{
		\begin{matrix}
			1 & \mbox{if } R_{OO}(t)<R_{OO}^{(c)} \\
			0 & \mbox{otherwise.}
		\end{matrix}
		\right.
	\end{matrix}
\end{equation}
Then, the quantity $\langle \tau_{HB} \rangle$ is evaluated by fitting the simulation results for $- dC_{HB}(t)/dt$ by Eq.~(\ref{eq: tau_HB_3}).

To characterize the kinetics of the HBs, it is necessary to define the \textit{local coordination number} $\mathcal{N}$ of a molecule. It determines number of the HBs, in which a molecule participates. Note that the quantity $\mathcal{N}$ is similar in its physical meaning to the first coordination number, but it is not the same, since it takes into account only those neighboring molecules that satisfy the geometric criterion of the HB. The average time for which a molecule is able to hold $\mathcal{N}$ bonds and thus maintain a given value $\mathcal{N}$ of the local coordination number defines the \textit{average coordination lifetime}~$\langle \tau_{\mathcal{N}} \rangle$.  The dynamics of the HB network occurs due to the formation of new HBs at each molecule and the breaking of existing bonds. Therefore, it seems reasonable to introduce the \textit{average waiting time} $\langle \tau_{i \to j} \rangle$, which characterizes the average time of continuous stay of a molecule in a state with $\mathcal{N}=i$ bonds before that molecule passes into a state with $\mathcal{N}=j$ bonds.  Then, the quantity $\langle \tau_{i \to j} \rangle^{-1}$ represents the frequency with which a molecule changes the $i$th coordination number to the $j$th coordination number. The values of the quantities $\mathcal{N}$, $\langle \tau_{\mathcal{N}} \rangle$ and  $\langle \tau_{i \to j} \rangle$ are estimated from direct analysis of the simulation data~\cite{AVM/GBN_JPCB_2013}.

\section{Results and Discussion}\label{sec: Results&Discussion}
	
\subsection{Liquid-liquid crossover}

In the LDL state, it is energetically favorable to form short- and medium-range order with directed bonds, the energy of which is comparable to the energy  $\varepsilon_{HB} \simeq 0.2$~eV of dimers of water molecules~\cite{HB_energy1,HB_energy2}. Consequently, the LL-crossover will occur at pressures that will brings the energy $E_p = p \Delta V_0$ into the local environment of water molecules comparable to the energy $(0.1 \div 1) \varepsilon_{HB}$; since no rigid bonds between molecules are formed as such. Here, $V_0 \simeq (4/3) \pi R_0^3$ is the volume per molecule, where $R_0 \simeq 3.0 \cdot 10^{-10}$~m is the distance between the centers of two hydrogen-bonded water molecules. It is reasonable to take the change in this volume as $\Delta V_0 \simeq 0.1 V_0$, then one obtains $\Delta V_0 \sim 1.0 \cdot 10^{-29}$~m$^3$. From this one finds that these pressures must be of the order of $10^3 \div 10^4$~atm, i.e., one gets the values that coincide in order with the actual pressures $p_{LL}$ of the observed LL-crossover [the thick red line in Figure~\ref{fig: phas_diagram}(a)]. In addition, it becomes clear from this point why this crossover is not observed at other, higher or lower pressures.

Specificity of the LL-crossover in water, related to the structural change, should certainly be reflected both in the dynamics of the water molecules and in the kinetics of the formation of the HBs. Molecular dynamics simulations using a given intermolecular interaction potential $U(\textbf{r})$ could be a suitable tool for this kind of study. All the results given below are derived from molecular dynamics simulations with the TIP4P/2005 potential~\cite{TIP4P-2005_1}.
	
\begin{figure}[h!]
		\includegraphics[scale=0.27]{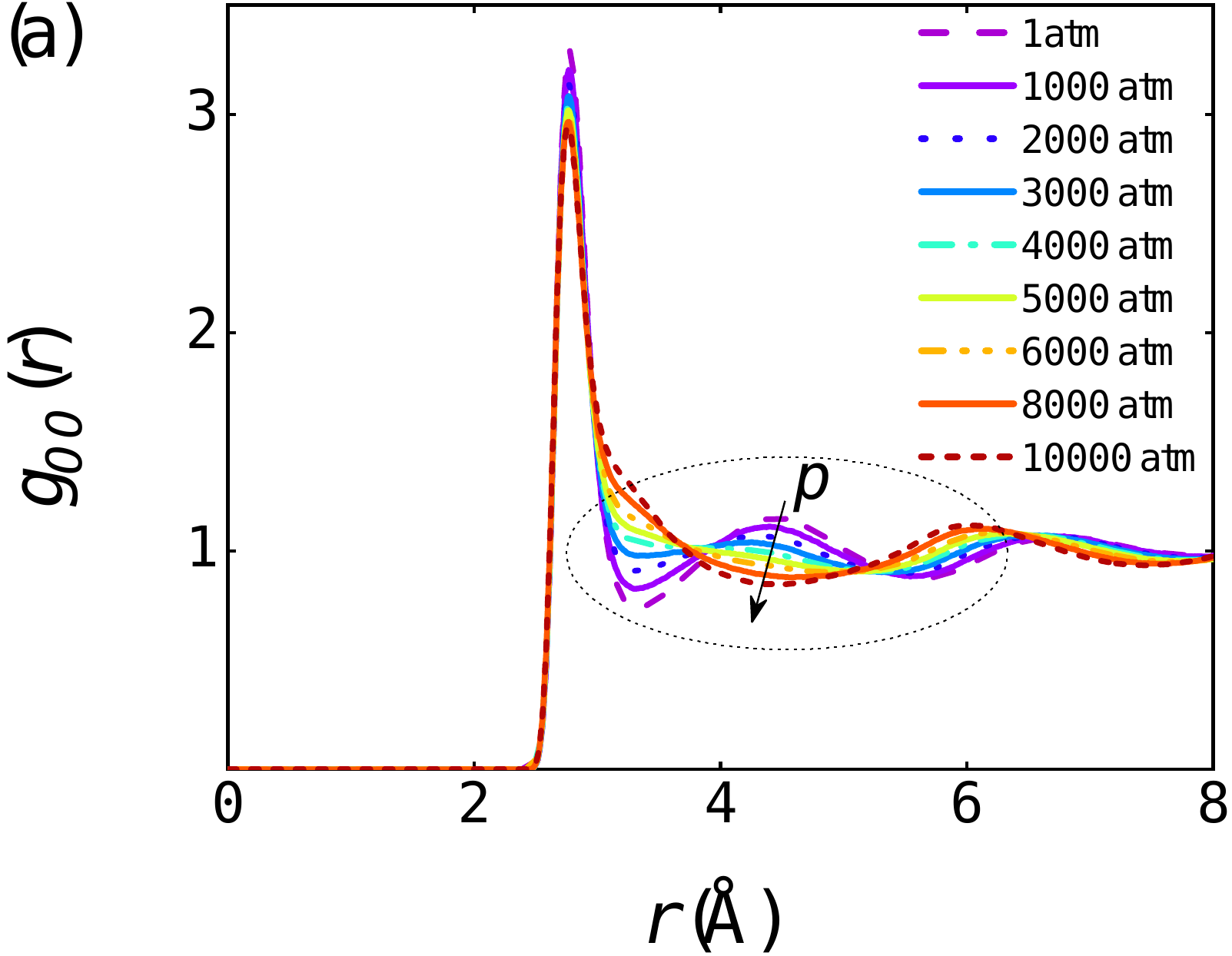}
		\hskip 0.9cm
		\includegraphics[scale=0.27]{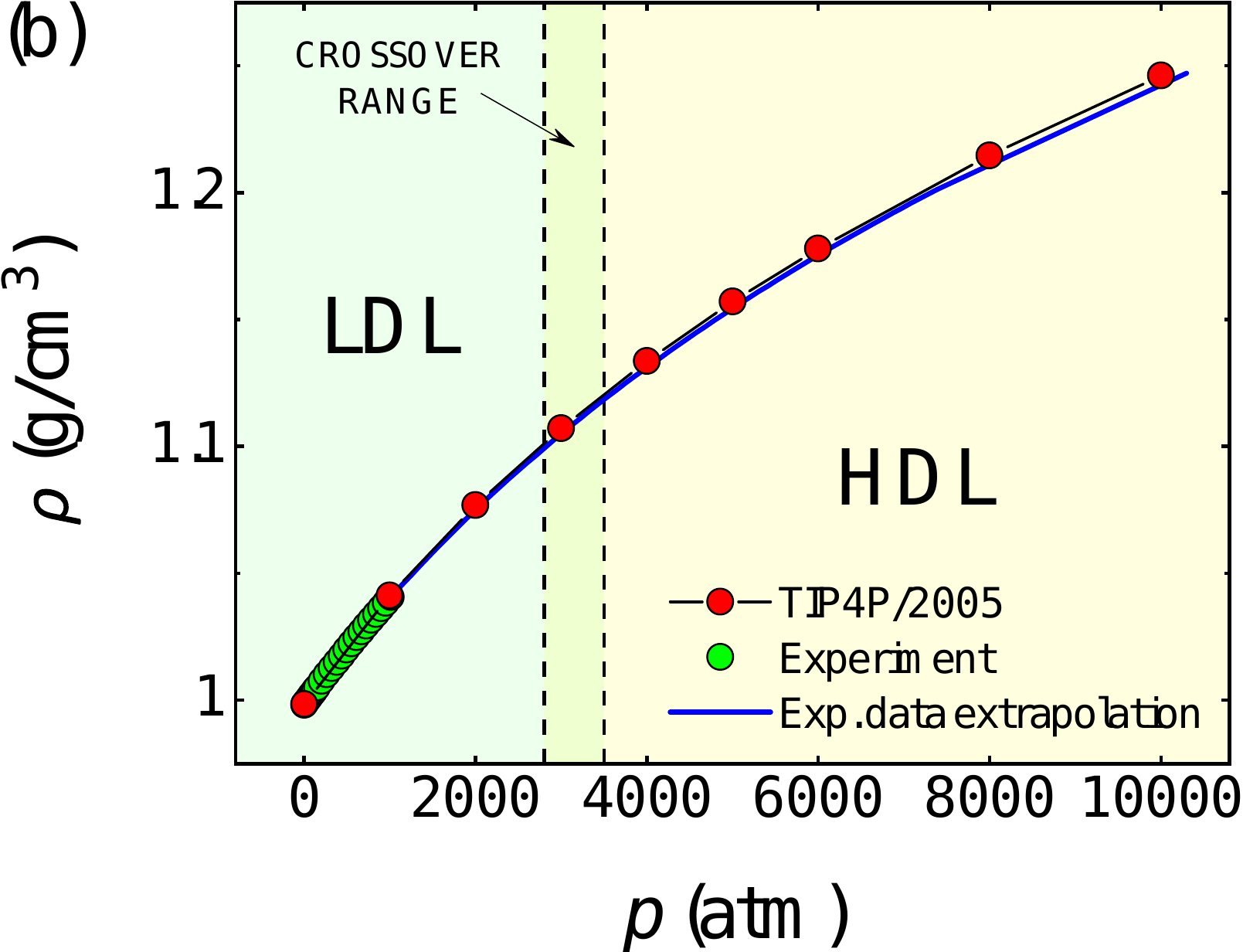}\\
		\vskip 0.3cm
		\includegraphics[scale=0.27]{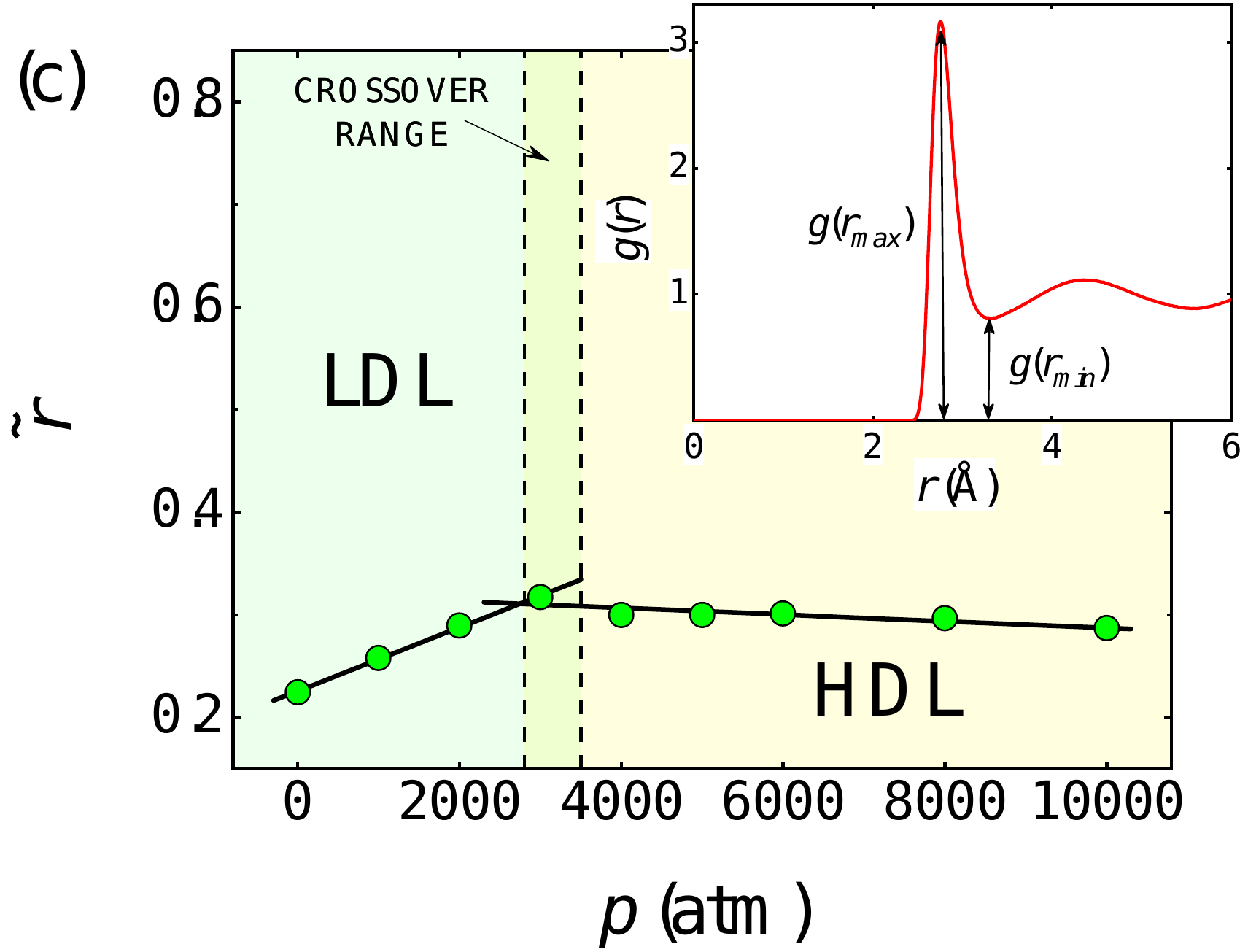}
		\hskip 0.5cm
		\includegraphics[scale=0.27]{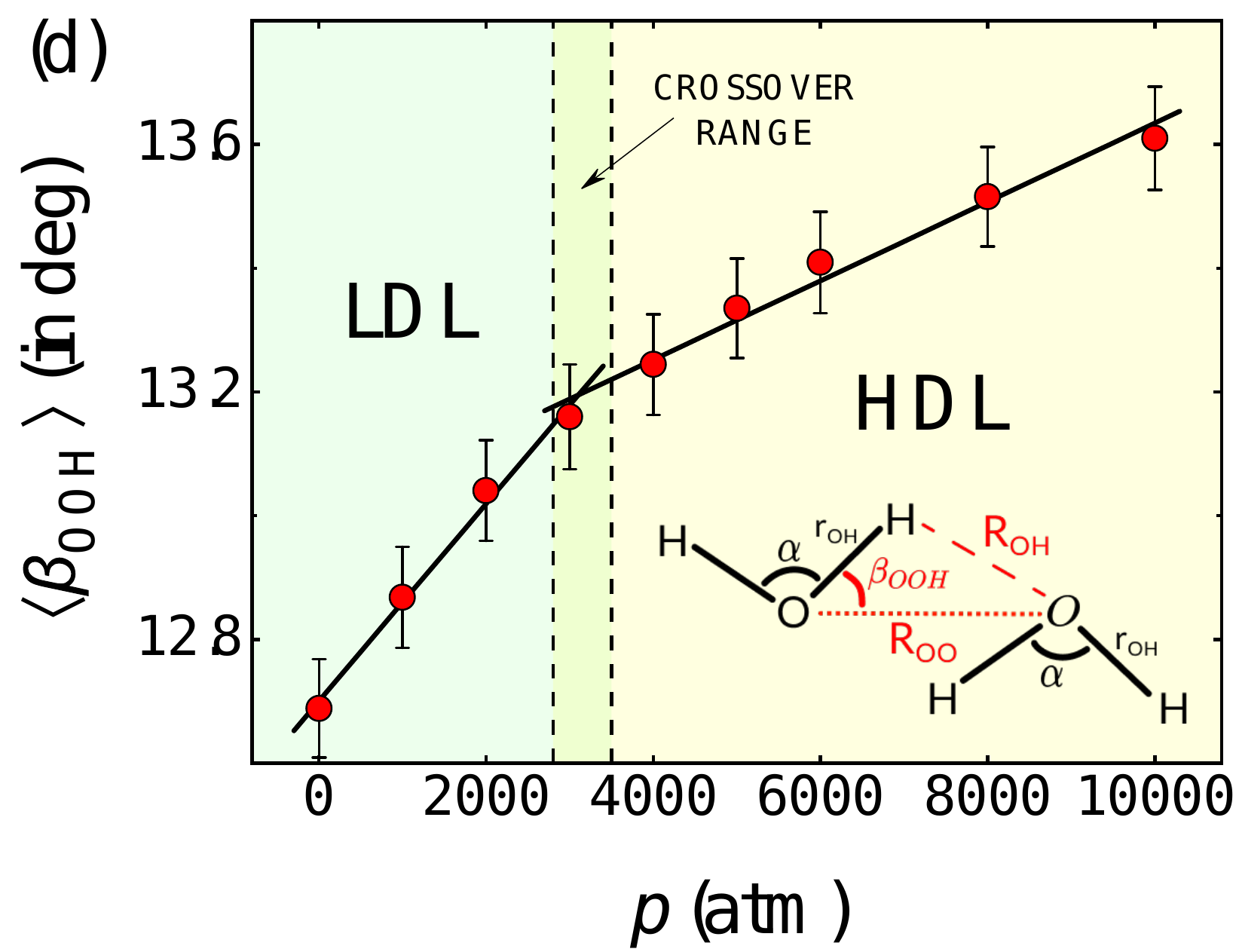}\\
		\vskip 0.3cm
		\includegraphics[scale=0.27]{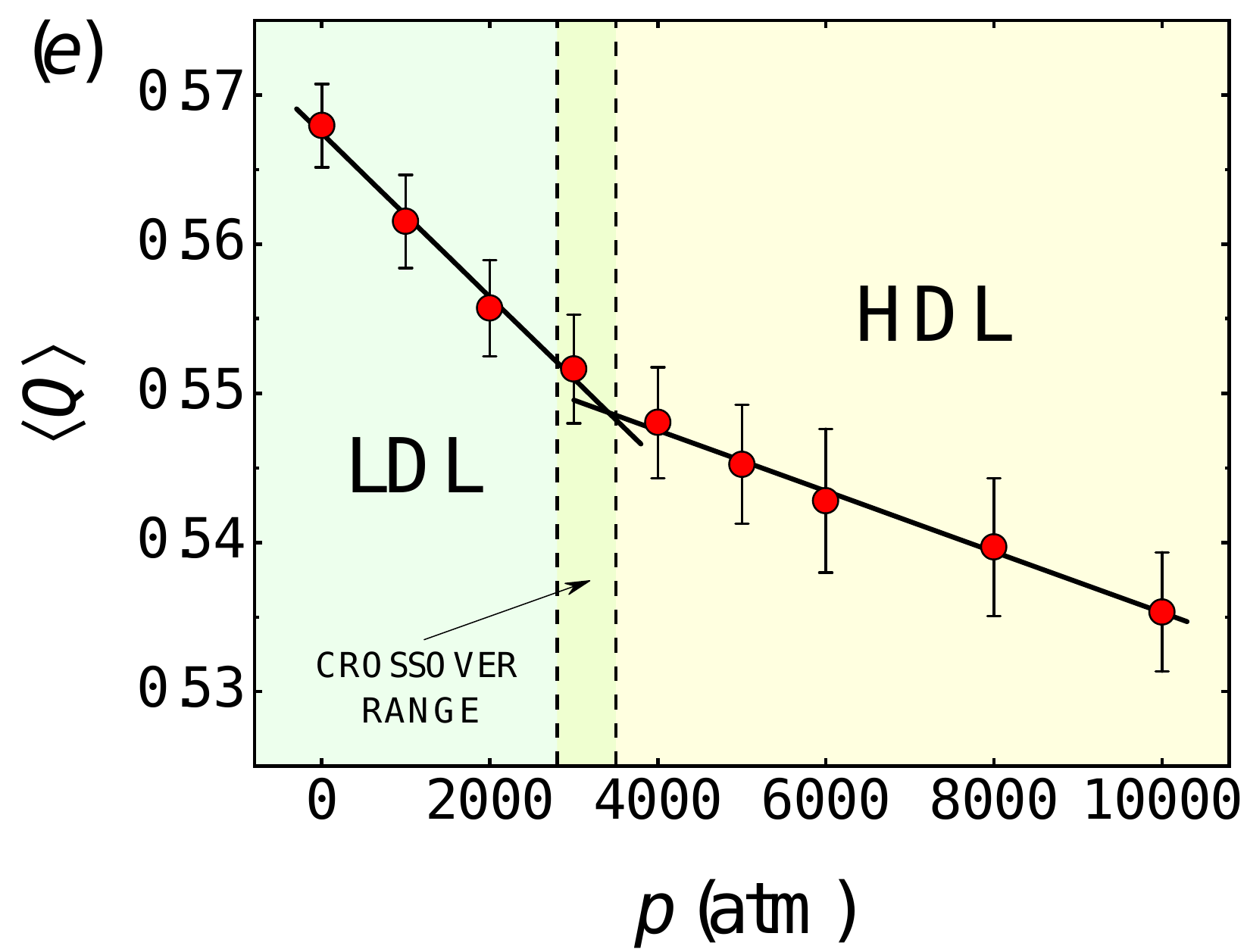}
		\hskip 0.8cm
		\includegraphics[scale=0.27]{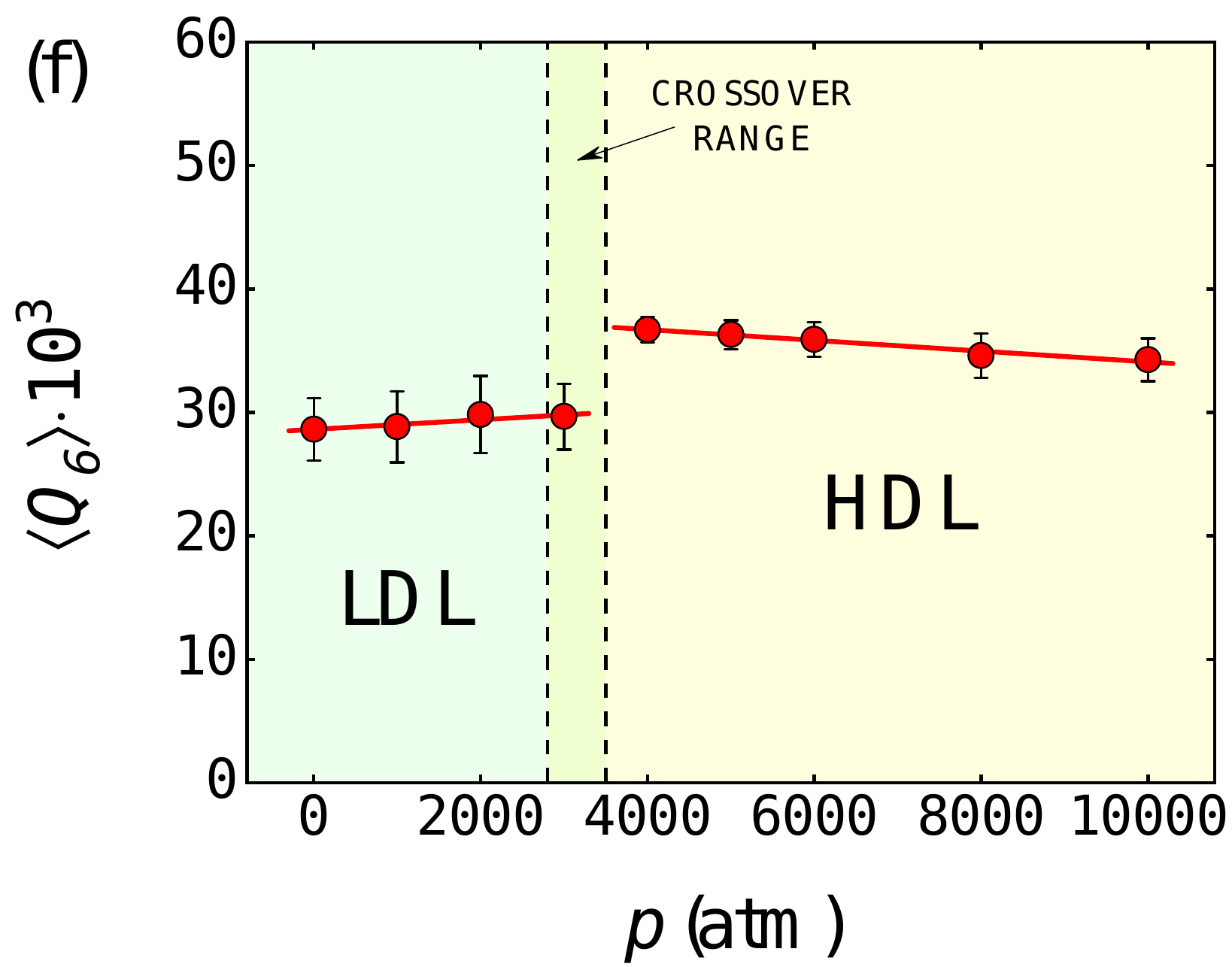}		
		
		\caption{
(Color online)  Structural characteristics calculated for equilibrium states of the water model at the temperature $T=293$~K and different pressures $p$:
(a) radial distribution function $g_{OO}(r)$;
(b) density as a function of the pressure $p$;
(c) Wendt-Abraham parameter~$\tilde{r}$;
(d) average HB angle $\langle \beta_{OOH} \rangle$;
(e)  tetrahedral order parameter $\langle Q \rangle$; and
(f) orientational order parameter $\langle Q_6 \rangle$.
}
		\label{fig: struct_parameters}
	\end{figure}

\textit{Structure.} --
If one considers the states of equilibrium liquid water along the isotherm $T=293$~K [see Figure~\ref{fig: phas_diagram}(b)], then in the pressure dependences of local structural characteristics the LL-crossover does appear at pressures in the vicinity of $p_{LL} \simeq 3\,150 \pm 350$~atm (see Figure~\ref{fig: struct_parameters}).
From the radial distribution function of the oxygen atoms $g_{OO}(r)$, which set the centers of mass of the water molecules, it follows that the second coordination sphere shifts to a smaller distance with increasing pressure and  collapses at the pressure $p_{LL}$ onto the first coordination sphere. It is noteworthy that for a macroscopic characteristic such as density, no peculiarities are observed over the entire pressure range covered. This can be seen in Figure 2(b), where the simulation results for the density $\rho(p)$ are compared with the available experimental data as well as with the results of the equation of state developed from the experimental data~\cite{Kell_Density_Exp,floriano2004dielectric}.
On the isotherm $T=293$~K, the significant changes of the average HB angle $\langle \beta_{OOH} \rangle$, the tetrahedral order parameter $\langle Q \rangle$, the Wendt-Abraham parameter $\tilde{r}$, the orientational order parameter $\langle Q_6 \rangle$ are revealed at the pressures associated with the pressure $p_{LL}$ [see Figure~\ref{fig: struct_parameters}(c--f)]~\cite{Errington_Q_order_parameter,Q6_parameter,Q6_parameter_AVM}.
In the crossover region, the orientational order parameter $\langle Q_6 \rangle$ shows a jump in values, although this jump is insignificant in magnitude. The other structural parameters behave continuously. This is consistent with some of the previous results for the LL-crossover obtained from molecular dynamics simulations~\cite{Khusnutdinoff_Mokshin_JNCS_2011,Khusnutdinoff_Mokshin_PhysA_2012,Yaping_PNAS_2013}.
	
	\begin{figure}[h!]
		\centering
		\includegraphics[scale=0.33]{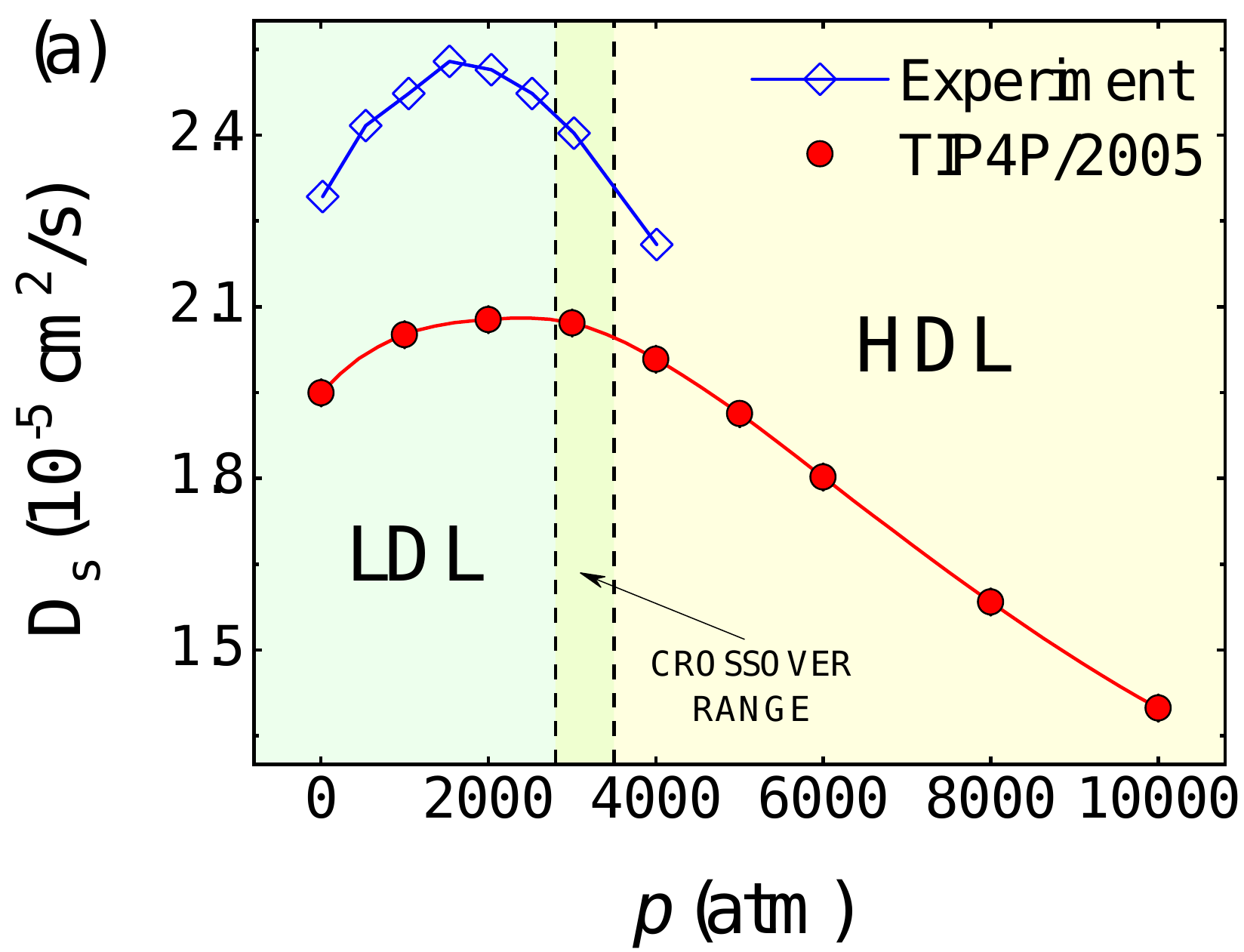}
		\vskip 0.5cm
		\includegraphics[scale=0.33]{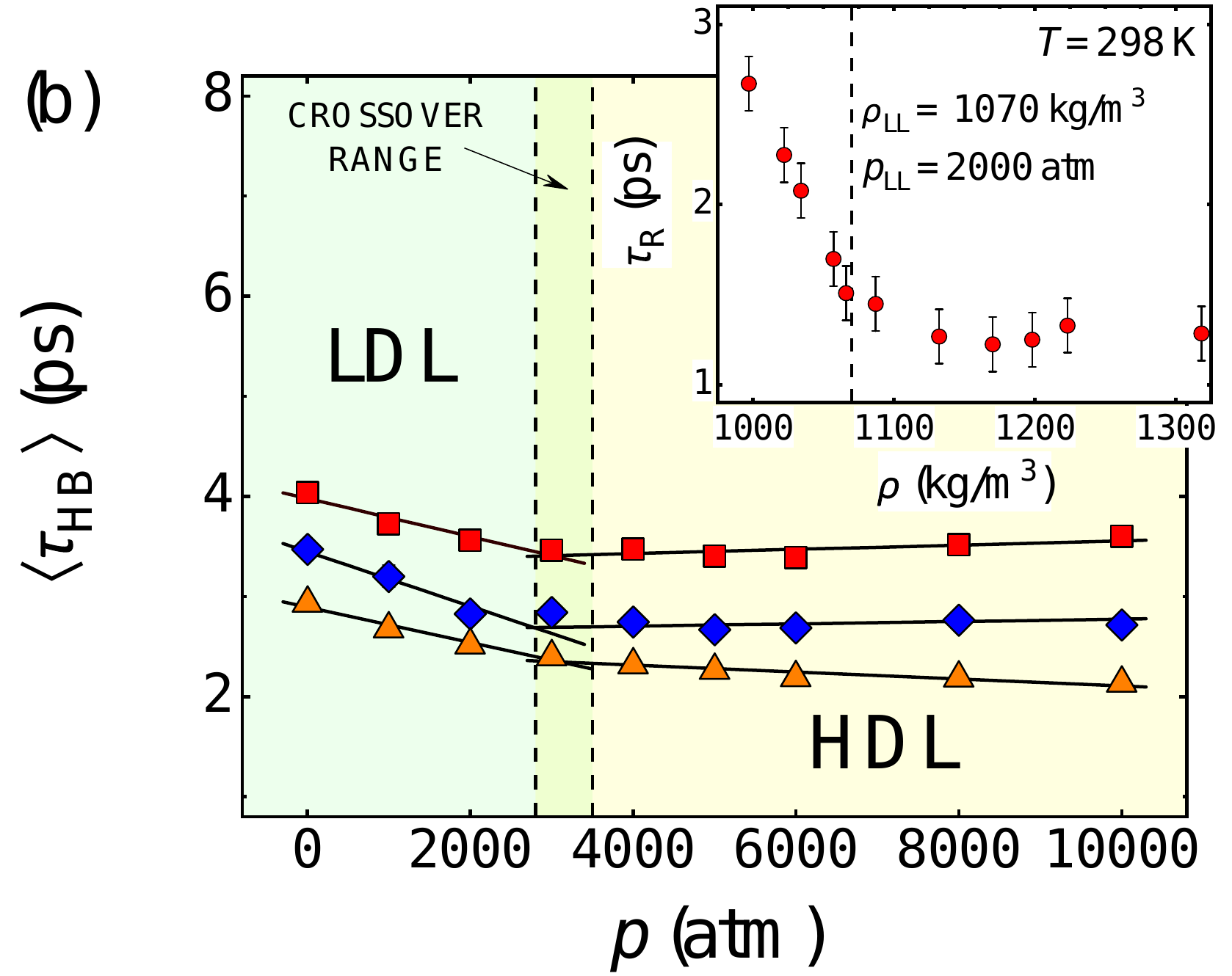}
		
		\caption{(Color online) (a) Self-diffusion coefficient $D_s$ for states at different pressures on the isobar $T=293$~K: simulation results and experimental data~\cite{self_diffusion_water}. (b) Main: Average HB lifetime $\langle \tau_{HB} \rangle$ calculated with different definitions:
		the orange triangles ($\blacktriangle$) represent direct estimates (see definition (a) in Sec. ``Methods'');
		the red squares ($\blacksquare$) represent results obtained with Eq.~(\ref{eq: tau_HB_2});
		the blue rhombuses ($\diamond$) correspond to the results obtained using Eq.~(\ref{eq: tau_HB_3}).
		Inset: Experimental rotational anisotropy time constant as a function of density at the temperature $T=298$~K (Ref.~\citenum{infrared_spectr}).}
		\label{fig: self_diff_HB}
	\end{figure}
	
\textit{Kinetics of hydrogen bonds.}~--
Significant changes in the dynamics of water molecules appear in the vicinity of the LL-crossover. In the case of the LDL states at pressures $p < p_{LL}$, the mobility of molecules is greater for states with higher pressures. For water, as a system with directional intermolecular bonds, this is quite expected. This is because at higher pressures the selected directions in the molecular interaction start to appear weaker, and the effective intermolecular interaction becomes more isotropic. As a result, at the pressures corresponding to the LDL states, the self-diffusion $D_s(p)$ as a function of the pressure $p$ increases, and the average lifetime $\langle \tau_{HB} \rangle$  of the HB with the pressure $p$ decreases (see Figure~\ref{fig: self_diff_HB}). In the HDL states, the anisotropy due to the characteristic water intermolecular interaction practically does not manifest itself. As a consequence, the physical characteristics as a function of the pressure should have a behavior similar to that observed in simple liquids. Then, the average lifetime $\langle \tau_{HB} \rangle $ of the HB takes the meaning of the characteristic neighborhood time of a pair of molecules, which is practically independent of the pressure $p$. In turn, the mobility of molecules should decrease as the system becomes more dense, as it is typical for simple liquids. This is manifested clearly in the self-diffusion coefficient $D_s$, the values of which decrease with increasing pressure $p$.  The above conclusions are completely supported by the results of ultrafast infrared pump-probe spectroscopy, which indicate that the rotational anisotropy of water molecules decreases with increasing pressure in the LDL state and that the rotational anisotropy almost completely disappears at the LL-crossover [see inset in Figure~\ref{fig: self_diff_HB}(b)]~\cite{infrared_spectr}.
	
\begin{figure}[h!]
		\centering
		\includegraphics[scale=0.35]{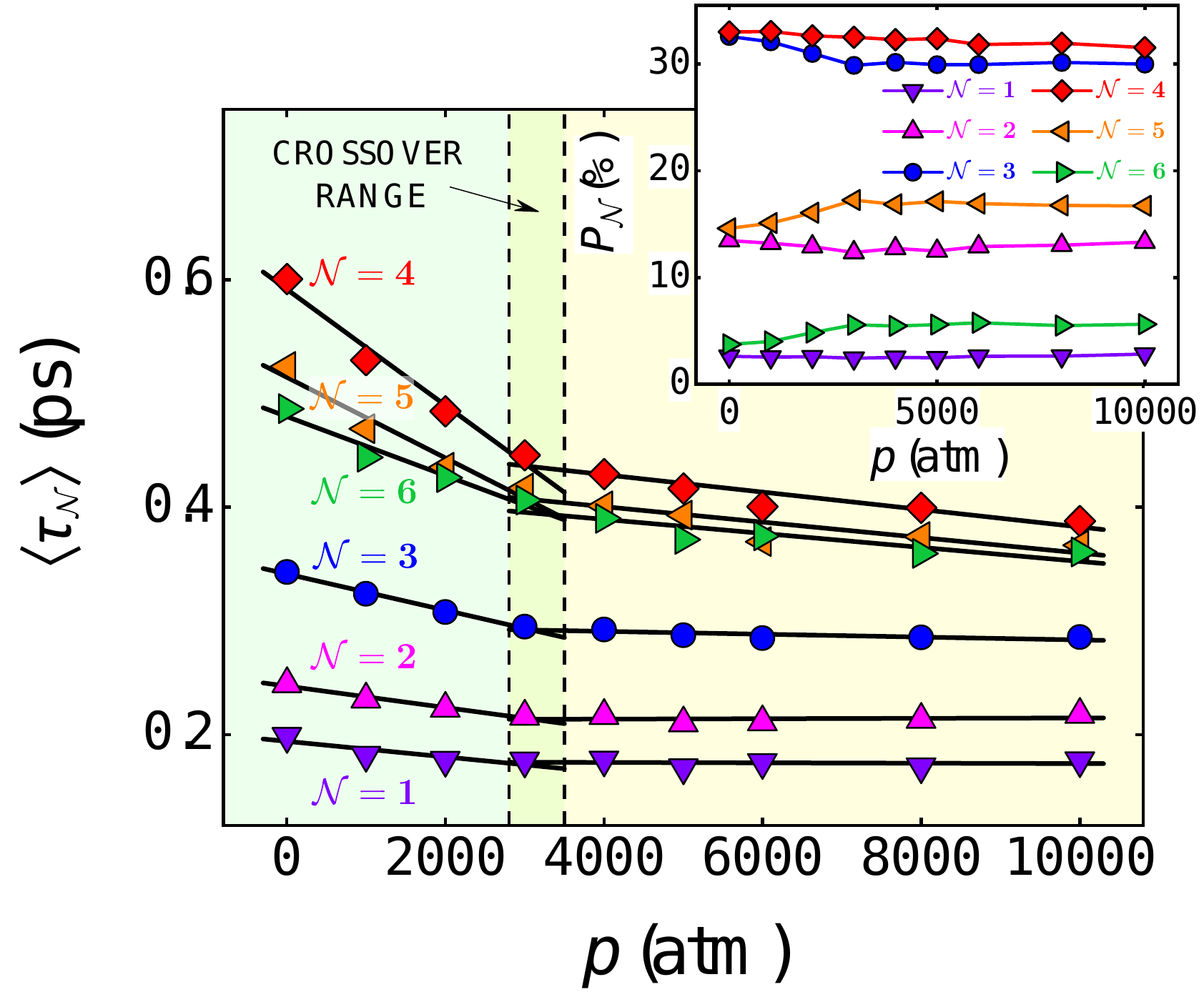}
		
		\caption{(Color online) Main: Average lifetimes $\langle \tau_\mathcal{N} \rangle$ of coordination numbers $\mathcal{N}=1,\; 2,\;\ldots,\;6$ at different pressures at the isotherm $T=293$~K. Inset: Occurrence probabilities $P_{\mathcal{N}}$ of the coordination number $\mathcal{N}$, where $\mathcal{N}=1,\; 2,\;\ldots,\;6$, for an arbitrary water molecule.}
		\label{fig: tau_N}
	\end{figure}

Since the LL-crossover is caused by changes in the local structure, it is useful to consider in detail the local coordination number $\mathcal{N}$ of molecules and the average coordination lifetime $\langle \tau_\mathcal{N} \rangle$.  By its physical nature, a water molecule is four-coordinated~\cite{coordination_1,coordination_2}, i.e., $\mathcal{N}=4$, where two bonds can belong to the negative charge concentration region of a molecule and two bonds can belong to two positively charged regions. In the case of liquid phase, the number of bonds $\mathcal{N}$ per molecule varies with time and may be more or less than four, since each of the charge regions of an arbitrary molecule forms a field of central forces. In fact, the local coordination numbers $\mathcal{N}=3$ and $\mathcal{N}=4$ are realized with equal probability $\sim 33$\;\%  in water (see inset in Figure~\ref{fig: tau_N}). The numbers $\mathcal{N}=2$ and $\mathcal{N}=5$ occur with equal probability $\sim 15$\;\%, and the numbers $\mathcal{N}=1$ and $\mathcal{N}=6$ occur with probability $\sim 2$\;\%. At the same time, these probabilities do not vary under the LL-crossover.
	
The average coordination lifetimes of molecules, $\langle \tau_1 \rangle$, $\langle \tau_2 \rangle$, $\dots$, $\langle \tau_6 \rangle$, decrease with increasing the pressure ~$p$ (see Figure~\ref{fig: tau_N}), and at the LL-crossover the character of dependences of the quantities $\langle \tau_\mathcal{N} \rangle$ on the pressure $p$ changes in a similar way as for the average HB lifetime $\langle \tau_{HB} \rangle$ [Figure~\ref{fig: self_diff_HB}(b)]. The most stable local configurations are those, where water molecules have a local coordination number $\mathcal{N}=4$, and the lifetime $\langle \tau_4 \rangle$ of such the configurations are the longest. It is noteworthy that high-density configurations with the local coordination numbers $\mathcal{N}=5$ and $6$ turn out to be of higher priority and are characterized by longer lifetimes than low-density ones with $\mathcal{N}=3$, $2$ and $1$, that is to be expected, when a molecular system with directed intermolecular bonds is in a high-density disordered state. Here, a regular network of HBs between water molecules is not formed, as, for example, in the case of crystalline ice.
	
\subsection{Free energy landscape}
	
\begin{figure}[h!]
		\centering
		\includegraphics[scale=0.18]{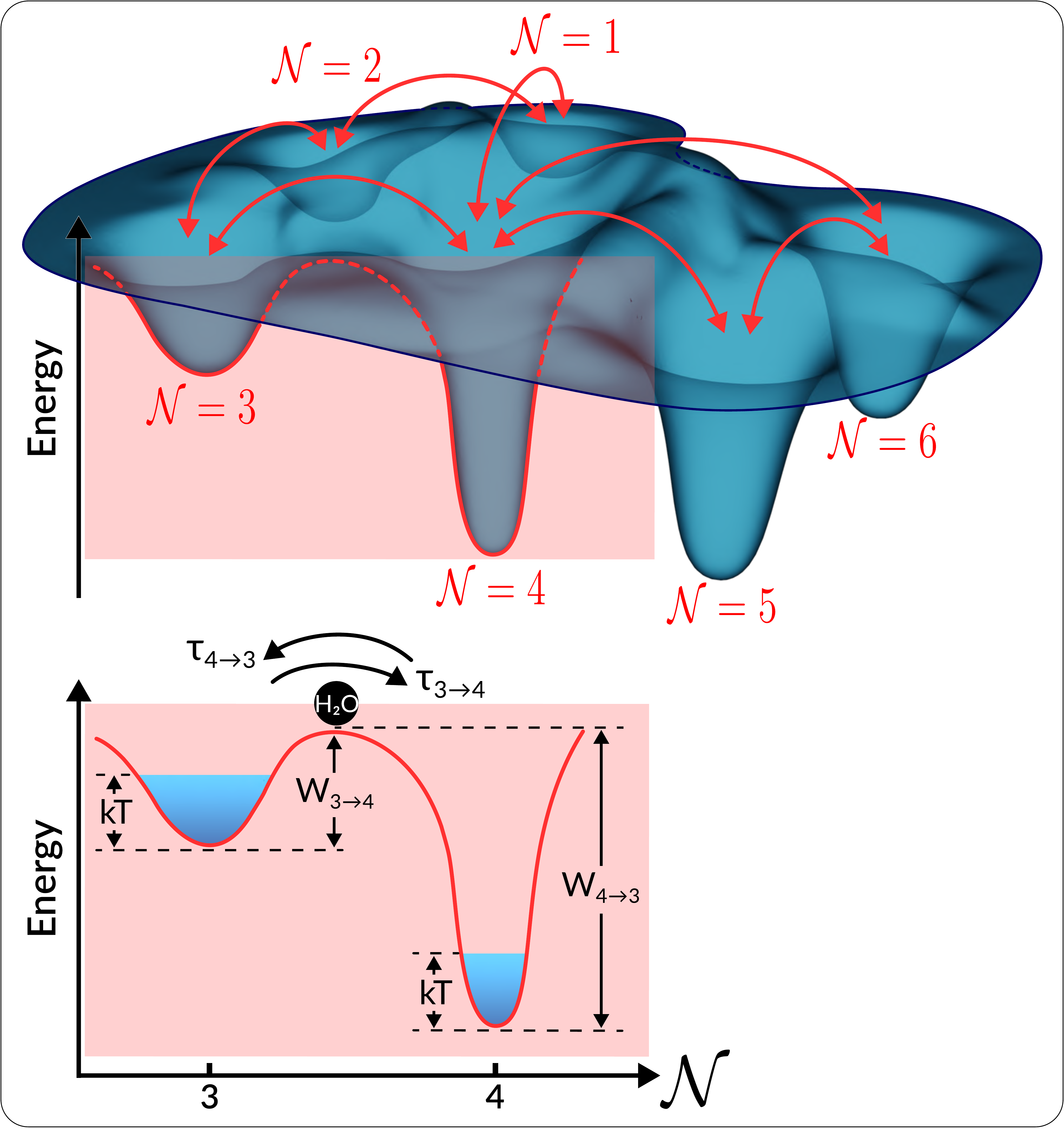}
		
		\caption{(Color online) Topological surface representing the free energy landscape $E(\overline{\mathcal{N}})$, where $\mathcal{N}$ is a local coordination number taking values $0$, $1$, $2$, $\dots$, $6$. The real dynamics of an arbitrary water molecule corresponds to the motion along this landscape with falling in the minima.}
		\label{fig: energy_surface}
	\end{figure}

\begin{figure}[t!]
		\centering
		\includegraphics[scale=0.33]{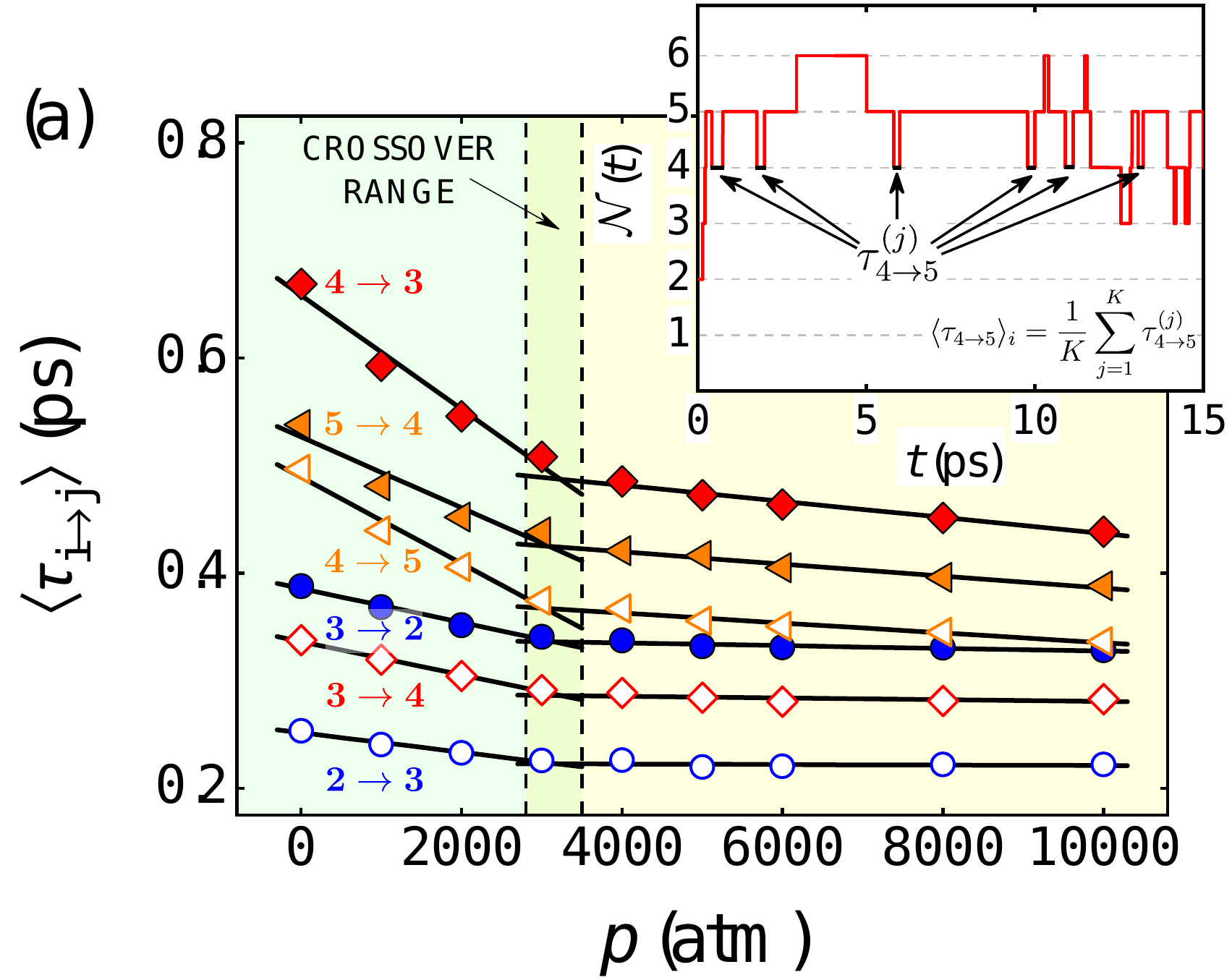}
		\vskip 0.5cm
		\includegraphics[scale=0.33]{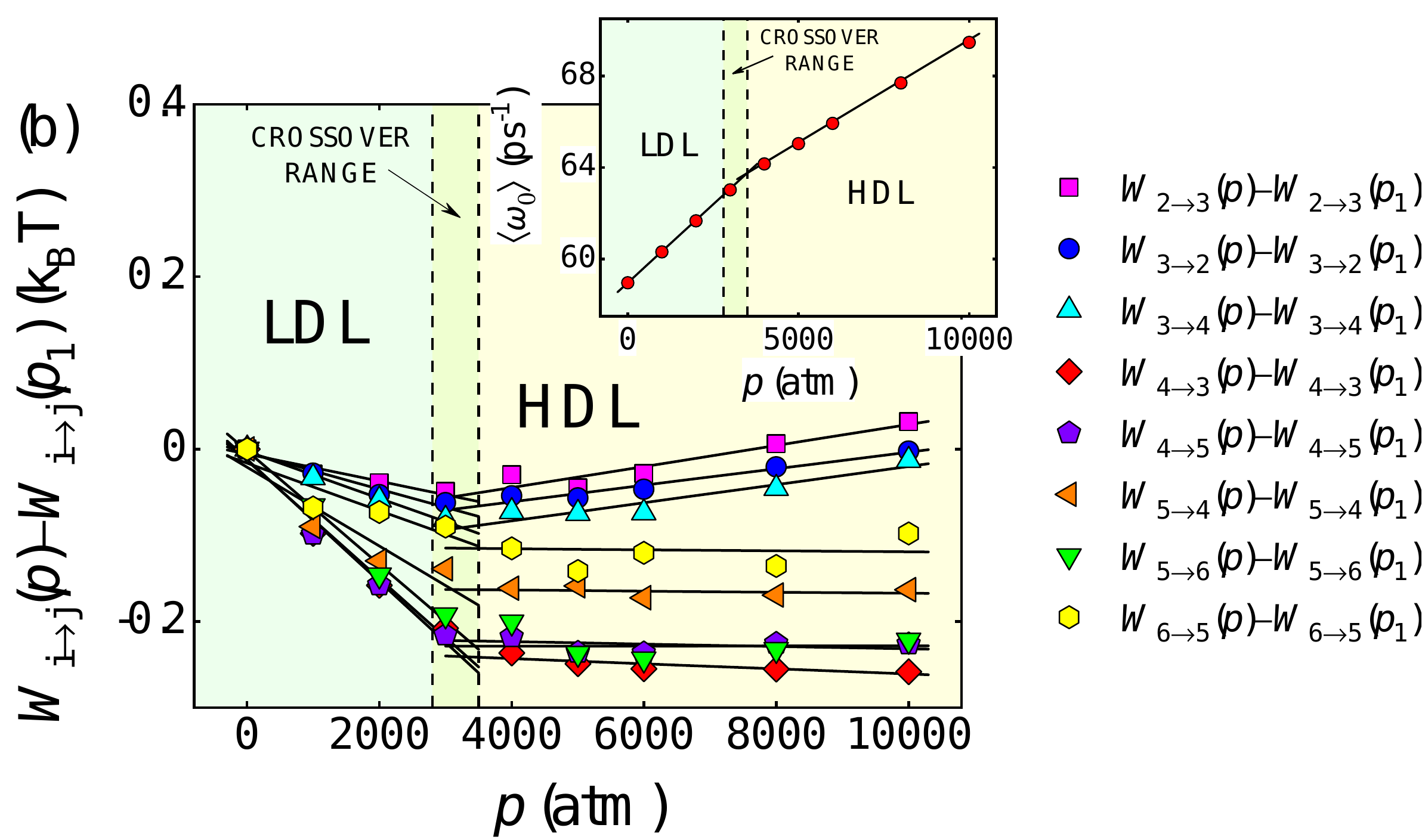}
		
		\caption{(Color online) (a) Main: Average waiting times $\langle \tau_{i \to j} \rangle$ for a molecule to transition from a state with a coordination number $\mathcal{N}=i$ to a state with a number $\mathcal{N}=j$ at different pressures $p$, where $i,\,j=2,\;3,\;4,\;5$. Inset: Changing coordination number $\mathcal{N}$ of an arbitrary $i$th water molecule over time $t$. The figure explains how the average waiting times $\langle \tau_{i \to j} \rangle$ are determined using the example case of $\langle \tau_{4 \to 5} \rangle$.
			(b) Main: Change of transition activation energies $W_{i\to j}(p)$ relative to their values for the state at pressure $p_1=1$~atm, i.e. $W_{i\to j}(p=1~\mathrm{atm}.)$.
			Inset:~The average vibration frequency~$\langle \omega_0 \rangle$ of water molecules for states at different pressures $p$.}
		\label{fig: tau_ij}
	\end{figure}

It is convenient to provide an interpretation of the HB kinetics by means of the \textit{free energy landscape} $E(\overline{\mathcal{N}})$ in the abstract space of local coordination numbers $\overline{\mathcal{N}}=\{0,\;1,\;2,\; \ldots \}$ (see Figure~\ref{fig: energy_surface}).
The minima of this landscape will correspond to certain values of $\mathcal{N}$, and the dynamics of an arbitrary water molecule will correspond to movement along the landscape $E(\overline{\mathcal{N}})$. Obviously, a shape of the landscape (depths of different minima, barriers) is determined by thermodynamic state of a system. The more stable the local configuration with a given coordination number, the deeper the corresponding minimum will be. A graphical explanation is given in schematic Figure~\ref{fig: energy_surface}. The transition from one minimum with $\mathcal{N}=i$ to another with $\mathcal{N}=j$ is characterized by a certain transition probability and the average waiting time $\langle \tau_{i \to j} \rangle$. In turn, the quantities $\langle \tau_{i \to j} \rangle$ are related to the average coordination lifetimes $\langle \tau_\mathcal{N} \rangle$ as follows:
	\begin{eqnarray}
		\langle \tau_\mathcal{N} \rangle &\equiv& \langle \tau_i \rangle =  \sum_{j=0}^{6} P_{i \to j} \, \langle \tau_{i \to j} \rangle, \\
		i,\;j&=&0,\,1,\,2,\;\ldots,\,6; \nonumber\\
		i &\neq& j, \nonumber
	\end{eqnarray}
where $P_{i \to j}$ is the probability of changing the $i$th coordination number to the $j$th one.
The times $\langle \tau_{i \to j} \rangle$ are pressure-dependent: they decrease linearly with increasing pressure values, showing changes in the LL-crossover region. Highly coordinated molecular states with $\mathcal{N}=4$ and $5$ appear to be most stable before transitions to the states with lower coordination numbers $\mathcal{N}=3$ and $4$, respectively [Figure~\ref{fig: tau_ij}(a)].

The frequencies  $\langle \tau_{i \to j} \rangle^{-1}$ of coordination number changes obey the following equation~\cite{TST}:
	\begin{eqnarray}
		\langle \tau_{i \to j} \rangle^{-1} &\sim& \langle \omega_0 \rangle \, \exp \left ( - \frac{W_{i \to j}}{k_B T} \right ), \\  i,\;j&=&0,\;1,\;2,\;\ldots,\,6; \nonumber \\
		i & \neq & j. \nonumber
	\end{eqnarray}
	Here, $W_{i \to j}$ is the free energy barrier for the transition from the minimum with $\mathcal{N}=i$ to the minimum with $\mathcal{N}=j$, and the quantity $\langle \omega_0 \rangle$ is the average frequency of vibrations of water molecules, which can be determined through the ratio of the first two frequency moments of the vibrational density of states $f(\omega)$ of water molecules:
	\begin{equation}
		\langle \omega_0 \rangle = \frac{\int \omega f(\omega) d\omega}{\int f(\omega) d\omega}.
	\end{equation}
	\noindent Note that the vibrational density of states $f(\omega)$ is defined here as the spectral density of the velocity autocorrelation function of molecules. The obtained results for $f(\omega)$  at different pressures $p$ are shown in Figure S1. The oscillations of an arbitrary molecule are determined by a size of the region (cell) formed by neighboring molecules. Therefore, it is quite natural that the frequency $\langle \omega_0 \rangle$ grows with density and increases linearly as a function of pressure:
	\begin{equation}
		\langle \omega_0 \rangle \propto \alpha \, p,
		\label{eq: omega_0}
	\end{equation}
	revealing changes in the LL-crossover region. Thus, one finds $\alpha = 13.5 \cdot 10^{3}$ and $8.8 \cdot 10^{3}$~m$^3$/(J$\cdot$sec) for the LDL and HDL states, respectively [inset in Figure~\ref{fig: tau_ij}(b)].

If one moves along the isotherm and considers equilibrium thermodynamic states at different pressures, it appears that the general shape of the free energy landscape $E(\overline{\mathcal{N}})$ persists. With increasing pressure, the depths of all minima in this landscape increase \textit{commensurately}. Thus, the baric dependences of the free energy barriers $W_{i \to j}(p)$ are reproduced by linear functions, and the character of these functions significantly changes at the LL-crossover [see Figure~\ref{fig: tau_ij}(b)], obeying the following general relation
	\begin{equation}
		\frac{dW_{i \to j}(p)}{dp}=\left\{
		\begin{array}{rl}
			 \Delta v_{i \to j}^{LDL}, &\quad{\rm for \ LDL \ states} \\
			 \Delta v_{i \to j}^{HDL}, &\quad{\rm for \ HDL \ states} \\
		\end{array}
		\right.
		\label{eq: barrier}
	\end{equation}
	where
	\[
	|\Delta v_{i \to j}^{LDL}| > |\Delta v_{i \to j}^{HDL}|.
	\]
The volume $v_{i\to j}$ means the magnitude of the changes of short-range order, when a molecule changes its coordination number from $\mathcal{N} = i$ to $\mathcal{N} = j$ in the corresponding LDL or HDL state. Then, the quantity $(p\, v_{i \to j})$ has a physical meaning of work, which is performed by a system, when a molecule changes its local environment with the coordination number $\mathcal{N}=i$ to $\mathcal{N}=j$, while $(p\, \Delta v_{i \to j})$ indicates the magnitude of the change in this work with unit pressure change. Positive values of $\Delta v_{i \to j}$ indicate that the local volume change $v_{i \to j}$ will be larger in a higher pressure state compared to the local volume change in a lower pressure state. In turn, negative values of $\Delta v_{i \to j}$ indicate a decrease of the local volume change $v_{i \to j}$ with increasing pressure. For the isotherm $T=293$~K, we have $\Delta v_{i \to j}^{LDL} \in [-2.9;\;-0.6]$~\AA$^{3}$ and $\Delta v_{i \to j}^{HDL} \in [-0.6;\;0.5]$~\AA$^{3}$.

\section{Conclusion}\label{sec: Conclusion}
	
The main results can be summarized as follows.

(i) For the water model on the isotherm $T=293$~K, the structural changes are found at the pressures $p_{LL}= 3\,150 \pm  350$~atm. The character of these changes in the structure is similar to that observed at the liquid-liquid first-order phase transition. However, in contrast to this phase transition, the observed structural changes occur smoothly,  typical of the liquid-liquid crossover, and are caused by the broadening of the first coordination shell due to changes in the second shell.

(ii) The self-diffusion is a non-monotonic function of pressure and attains a maximum in the neighborhood of the LL-crossover. In the region of LDL states, the weakening of the anisotropy in the interparticle interaction with pressure has an effect on the increase in the mobility of molecules and their self-diffusion. For HDL states, the self-diffusion decreases as the density of the system increases, which is due to the fact that anisotropy in the water intermolecular interaction practically does not manifest itself and that is typical for simple liquids.

(iii) Changes in the structure directly affect the kinetics of hydrogen bond network formation. It is found that the average hydrogen bond lifetime as well as the average lifetime of different coordination numbers decreases with increasing pressure, and the changes are detected at the LL-crossover. Furthermore, the average lifetimes of the coordination numbers are fractions of picoseconds, that is comparable to the characteristic time scale of self-diffusion of the molecules; and the stable long-lived hydrogen bonds in water are not formed even in the range of the HDL states.

(iv) The concept of the free energy landscape in the space of possible coordination numbers is proposed to describe hydrogen bonding kinetics. As found, with increasing pressure, the depths of all minima in this landscape increase commensurately. Free energy barriers for the transitions between the states with various coordination numbers as  functions of pressure are reproduced by the linear functions, and the slopes of these functions change significantly at the LL-crossover.

In addition, the obtained results lead to the following general conclusions related to the \textit{necessary condition} for the existence of the LDL/HDL transition in the system. The LLT as well as the LL-crossover are induced by pressure and occur in the systems with a specific interparticle interaction. It can be an interaction with pronounced anisotropy, where the non-sphericity of a potential is due to the presence of selected directions~\cite{network_forming_liquids} (as, for example, in water) or is due to the presence of some range of lengths corresponding to possible values of equilibrium interparticle distances (as, for example, in polyvalent metal melts~\cite{AVM_Gallium_2019}). Alternatively, it could be an isotropic interparticle interaction reproduced by a spherical-type potential, which must have a negative curvature region at distances smaller than the effective equilibrium interparticle distance. Due to these features of the potential at finite pressures, there appears a new correlation length characterizing an average effective particle size in the high-density state.

\begin{suppinfo}
	
	 Vibrational density of states (PDF).
	
\end{suppinfo}

\section*{Declaration of Competing Interest}
	
\noindent The authors declare that they have no known competing financial interests or personal relationships that could have appeared to influence the work reported in this paper.

\begin{acknowledgement}
		
\noindent The authors are grateful to V. V. Brazhkin, V. N. Ryzhov and D. L. Melnikova for useful comments.
The work was supported by the the Kazan Federal University Strategic Academic Leadership Program (PRIORITY-2030).

		
\end{acknowledgement}

\bibliography{paper}
	
\end{document}